\documentclass[aps,prl,superscriptaddress,showpacs,twocolumn]{revtex4-1}
\usepackage{amssymb}
\usepackage{graphicx}
\usepackage{dcolumn}
\usepackage{bm}
\usepackage{amsmath}

\usepackage[usenames]{color}
\usepackage{textcomp}
\begin{document}
\title{Heteronuclear magnetisms with ultracold spinor bosonic gases in optical lattices}
\author{Yongqiang Li}
\email{li\_yq@nudt.edu.cn}
\affiliation{Department of Physics, National University of Defense Technology, Changsha 410073, P. R. China}
\author{Chengkun Xing}
\affiliation{Key Lab of Quantum Information, CAS, University of Science and Technology of China, Hefei, 230026, P.R. China}
\author{Ming Gong}
\email{gongm@ustc.edu.cn}
\affiliation{Key Lab of Quantum Information, CAS, University of Science and Technology of China, Hefei, 230026, P.R. China}
\affiliation{Synergetic Innovation Center of Quantum Information and Quantum Physics, University of Science and Technology of China, Hefei, 230026, P.R. China}
\author{Guangcan Guo}
\affiliation{Key Lab of Quantum Information, CAS, University of Science and Technology of China, Hefei, 230026, P.R. China}
\affiliation{Synergetic Innovation Center of Quantum Information and Quantum Physics, University of Science and Technology of China, Hefei, 230026, P.R. China}
\author{Jianmin Yuan}
\affiliation{Department of Physics, National University of Defense Technology, Changsha 410073, P. R. China}
\affiliation{Department of Physics, Graduate School of China Academy of Engineering Physics, Beijing 100193, P. R. China}

\date{\today}

\begin{abstract}
    Motivated by recent realizations of spin-1 NaRb mixtures in the experiments, here we investigate heteronuclear magnetism in the Mott-insulating regime. Different from the identical mixtures where the boson (fermion) statistics only admits even (odd) parity states from angular momentum composition, for heteronuclear atoms in principle all angular momentum states are allowed, which can give rise to new magnetic phases. Various magnetic phases can be developed over these degenerate spaces, however, the concrete symmetry breaking phases depend not only on the degree of degeneracy, but also the competitions from many-body interactions. We unveil these rich phases using the bosonic dynamical mean-field theory approach. These phases are characterized by various orders, including spontaneous magnetization order, spin magnitude order, singlet pairing order and nematic order, which may coexist, especially in the regime with odd parity. Finally we address the possible parameter regimes for observing these spin-ordered Mott phases.
\end{abstract}


\maketitle
Ultracold atoms in optical lattices provide an unique and versatile platform for simulating interesting models in condensed matter physics and quantum optics~\cite{Lewenstein07}, including the Bose-Hubbard (BH) model~\cite{Greiner02}, Fermi-Hubbard model~\cite{Parsons16, Boll16}, Dicke model~\cite{Baumann10}, topological Haldane model~\cite{Gregor14} and toric code model~\cite{Dai17} {\it etc}. These endeavors have greatly enrich our understanding of many-body physics, especially about their spatial correlations, fluctuations, topological transitions and even their non-equilibrium dynamics. By selecting two hyperfine states from a $2j+1$ manifold, where $j=k$ or $k+{1\over 2}$ ($k \in \mathbb{Z}$), it is possible to realize the ferromagnetic (FM) and antiferromagnetic Heisenberg models in the deep Mott-insulating regimes with bosonic or fermionic atoms (see recent evidences \cite{Parsons16, Boll16}), which are cornerstone for magnetic phases in solid materials. Going beyond this general scenario to the regime with strong interactions, large spins and even long-range interactions~\cite{Krutitsky16} are possible with ultracold atoms~\cite{waguchi12,Capponi16,Stamper13}, which will exhibit intriguing features rarely seen or hardly accessible in solid materials.

\begin{figure}
    \centering
    \includegraphics[width=3.3in]{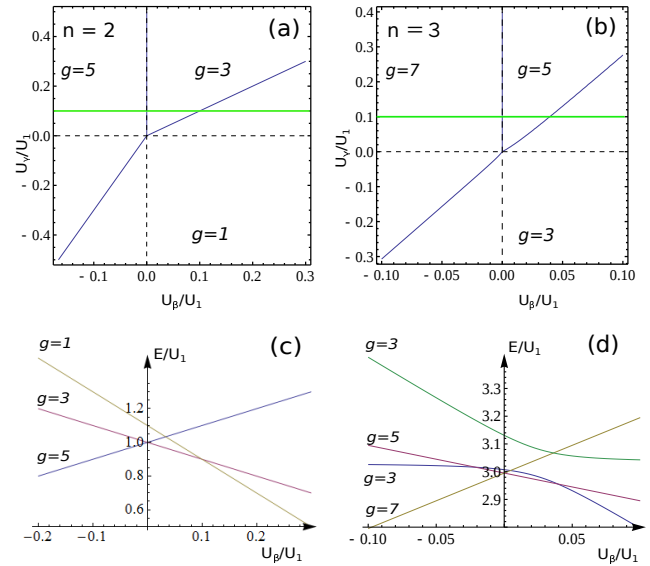}
    \vspace{-3mm}
    \caption{(Color online) Parities and degree of degeneracy for two (left) heteronuclear and three (right) heteronuclear atoms in a single well. Figures (a),(b) show the corresponding phase diagrams, and (c),(d) typical eigenvalues as a function of $U_\beta$ for $U_\gamma/U_1 = 0.1$ [green lines in (a),(b)]. }
    \label{fig-fig1}
\end{figure}

In optical lattices, the large spin manifold can be realized either by atoms with large $j$ in alkaline-earth atoms or by considering two or more small spin identical atoms~\cite{Gorshkov10, Honerkamp, Stenger98, Miesner99}. For instance, two identical spin-1 bosons can form a composite spin-2 bosons from angular momentum composition. The corresponding ground-state space is five-fold degenerate (angular momentum $F=2$ with degree of degeneracy $g = 2F+1 = 5$) or singlet ($F = 0$ and $g = 1$) since only even parity states are allowed for bosonic statistics. By contrast, for two identical fermions, only the odd parity states are allowed due to Pauli exclusive principle, thus $F = 1$. In both cases, the rotational symmetry for identical particles ensures that the effective spin models should only allow the isotropic Heisenberg term (direct product of two spins) and their powers.

In this Letter we mainly focus on heteronuclear magnetism in the Mott lobes in a three dimensional (3D) optical lattice. The absence of identity restriction admits both even and odd parity quantum states, which can give rise to new magnetic phases. We are motivated by recent experimental realizations of NaRb heteronuclear atoms in Wang's group~\cite{DJWang15, DJWang15a, DJWang16}, in which collision induced spin exchange between heteronuclear atoms is observed. We investigate the ground-state spin structures using the bosonic dynamical mean-field theory (BDMFT) approach. We find that the spin structures are not merely determined by ground-state degeneracy, but also their many-body competitions, which induce different types of symmetry breaking magnetic phases. Various phases in the deep Mott-insulating regime are unveiled, including the spin-singlet insulator (SSI), nematic insulator (NI), cyclic (C) phase, and different types of FM phases. These phases are characterized by a unique order parameter or by coexisting of several different order parameters. We even find an intriguing paired FM (pFM1) phase with $F = 1$, in which the nematic, ferromagnetic and singlet pairing orders coexist. Finally we discuss the possible experimental observation of these magnetic phases in the whole Mott-insulating regime.

{\it Model and Degeneracy}. Under the single mode approximation~\cite{hetero, hetero1, hetero2, hetero3, hetero4}, the 3D lattice can be described by the following generalized BH model (see details in Ref.~\onlinecite{Supp}),
\begin{eqnarray}\label{Hamil}
    \hat{H} &=& -\sum_{\langle ij \rangle,m,\sigma} t_m (b^\dagger_{im\sigma}b_{jm\sigma} + {\rm H.c.}) - \sum_{i,m}\mu_m n_{i m} \nonumber \\
        &+& \sum_{i,m} \bigg[\frac12U_{ m} n_{i m}(n_{i m}-1) + \frac12U^\prime_{ m} ({\bf S}^2_{i m}-2n_{i m})  \nonumber\\
        &+&  U_\alpha n_{i1}n_{i2}  + U_\beta {\bf S}_{i1} \cdot {\bf S}_{i2} + \frac {1}{3} U_\gamma \Theta^\dagger_i \Theta_i\bigg],
\end{eqnarray}
where $b^\dagger_{im\sigma}$ ($b_{im\sigma}$) is the bosonic creation (annihilation) operator of hyperfine state $\sigma  = \{1,0, -1\}$ for species $m=1, 2$ at lattice site $i$, $n_{i m}=\sum_\sigma n_{im\sigma}$ with $n_{im\sigma}\equiv b^\dagger_{im\sigma}b_{im\sigma}$ being the number of particle, ${\bf S}_{i m}\equiv b^\dagger_{im\sigma}{\bf \boldsymbol{\Gamma}}_{\sigma\sigma^\prime}b_{im\sigma^\prime}$ is the total spin operator with $\boldsymbol{\Gamma}_{\sigma\sigma^\prime}$ being the spin matrices for spin-1, $\Theta^\dagger \equiv b^\dagger_{i11}b^\dagger_{i2-1} - b^\dagger_{i10}b^\dagger_{i20} + b^\dagger_{i1-1}b^\dagger_{i21}$, $\mu_{ m}$ denotes the chemical potential, and $t_{ m}$ denotes the hopping amplitude between nearest neighboring sites. The $U$-terms describe the many-body interactions, which are related to the on-site Wannier functions and in principle can be tuned independently in experiments. For example, in general $U^\prime_{ m}/U_{ m}\ll1$ and $U_{\beta,\gamma}/U_\alpha \ll 1$, but these ratios can be tuned via microwave~\cite{micro_resonance, micro_resonance1} or optical Feshbach resonances~\cite{optical_resonance, optical_resonance1, optical_resonance2, optical_resonance3, optical_resonance4, optical_resonance5, optical_resonance6, optical_resonance7}.
Notice that $U_\beta$ describes the interactions between heteronuclear atoms, and is essential for heteronuclear spin exchange during collision, as shown in Ref.~[\onlinecite{DJWang15, DJWang15a, DJWang16}].
We stress that this model possesses both features of Fermi-Hubbard model~\cite{two_orbital1, two_orbital2, two_orbital3} and spinor BH model~\cite{hetero_theory}, due to the allowed odd and even parities.

\begin{figure}
\centering
\includegraphics[trim = 1mm 0mm 0mm 0mm, clip=true, width=0.49\textwidth]{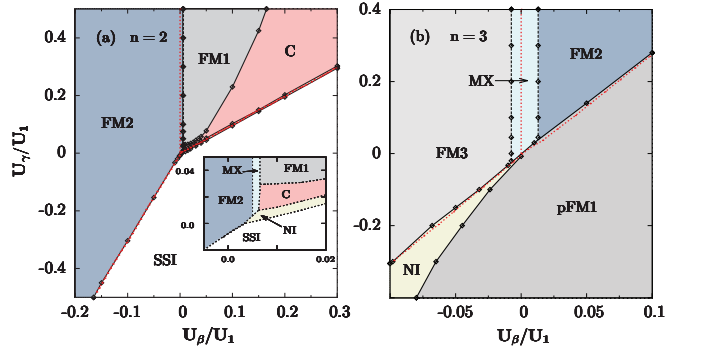}
\vspace{-7mm}
    \caption{(Color online) Phase diagrams for heteronuclear mixtures of spin-1 bosonic gases on a 3D cubic lattice in the typical Mott-insulating regime ($t_{1,2} = 0.01$, see Fig.~\ref{fig-fig4}) for $n = 2$ (a) and $3$ (b), respectively. Inset shows the zoom of the main figure near zero $U_\beta/U_1$ and $U_{\gamma}/U_1$. Other parameters are: $U_2/U_1$ = 1.92, $U_\alpha/U_1 = 1.0$, $U^\prime_1/U_1 = -0.005$ ($^{87}$Rb), and $U^\prime_2/U_2 = 0.037$ ($^{23}$Na). Notice that the red dashed lines are from Fig.~\ref{fig-fig1}.}
\label{fig-fig2}
\end{figure}

Magnetism is formed from the super-exchange interaction between the neighboring sites, which can induce direct coupling between all the quantum states in the degeneracy space. 
For this reason, it is instructive to firstly understand the ground-state degeneracy in each site by setting $t_{ m} = 0$. Our calculated results
are presented in Fig.~\ref{fig-fig1}. When each site contains two heteronuclear spin-1 atoms ($n = 2$), the angular momentum coupling rule for heteronuclear atoms allows all the possible angular momenta $F = 0, 1, 2$, with corresponding on-site degeneracy $g = 1, 3, 5$, respectively. The calculated phase diagrams are presented in Fig.~\ref{fig-fig1}(a) and (c), where the three boundaries are determined by
\begin{equation}\label{Eq_2}
    U_{\beta} = 0, \quad U_{\gamma} = U_\beta > 0, \quad U_{\gamma} = {1\over 3} U_\beta < 0.
\end{equation}
These boundaries are independent of $U_m'$. When three atoms ($n = 3$) are occupied in each site, the two identical bosonic atoms admit only even angular momenta, and then the angular coupling between between heteronuclear atoms yields $F = 1, 2, 3$ and $F = 1$, with corresponding $g = 3, 5, 7$ and $3$, respectively.  The phase boundaries in Fig.~\ref{fig-fig1}(b) are determined as,
\begin{eqnarray}
    &U_\beta   & = 0, \quad 3U_\beta = 4U_\gamma - \sqrt{\Delta} - 9U' > 0,  \text{  and }\nonumber \\
    &21U_\beta & = 4U_\gamma - \sqrt{\Delta} -9U' < 0,
\end{eqnarray}
where $\Delta = 81U'^2+81U_\beta^2 + 48U'U_\gamma + 16U_\gamma^2 - 6U_\beta (27 U' + 8U_\gamma)$ when $U' = U^\prime_1$. Different from the boundaries defined in Eq.~(\ref{Eq_2}), in this case the boundaries depend strongly on the values as well as the sign of $U_m'$, without which ($U' = 0$) the two equations collapse to a single line, $U_\gamma = {1\over 3} U_\beta$. The corresponding wave functions for these two cases are supplemented in Ref.~\onlinecite{Supp}.

The phase diagrams in Fig.~\ref{fig-fig1} are necessary, but not sufficient to understand the magnetic phases. In following we investigate these phases using the bosonic DMFT, which captures all the local quantum fluctuations exactly~\cite{BDMFT1,Hubener, Hubener1, Hubener2,
Hubener3, theory_boson3, BDMFT2, BDMFT3}. This method has been successfully applied to investigate the possible exotic magnetisms and superfluids in various models, including two-component spin-orbit coupled BH models~\cite{Liang15}, the spinor BH models~\cite{theory_boson3}. The reliability of this approach has been compared against the quantum Monte-Carlo simulations~\cite{QMC_boson}.
See more details about this approach in Ref.~\onlinecite{Supp}.

\begin{table}
    \caption{Characterization of different quantum phases for heteronuclear mixtures in an optical lattice. The definition of these orders ($\phi_{m\sigma}^1$, $\phi_{\alpha\beta}^2$, $\phi_p^3$, $M$ and $P$) can be found in the main text. The various magnetic orders are not measured (-) in the SF phase with $\phi^1_{m\sigma}\ne 0$.}
    \begin{tabular}{p{0.07\textwidth}|p{0.07\textwidth}p{0.07\textwidth}p{0.07\textwidth}p{0.07\textwidth}p{0.07\textwidth}} \hline \nonumber
    Phases         & $\phi^1_{m\sigma}$ & $\phi^2_{\alpha\beta}$    & $\phi^3_p$      & $M$                             &  $P$     \\\hline
    SF             & $\neq 0$            &  -                        & -               & -                               &  -        \\
    FM             & $=0$                &  $\ne 0$                  & $=0$            &     $\ne 0 $                    &  $\ne 0$   \\
    pFM            & $=0$                &  $\neq 0$                 & $\neq 0$        &     $\neq 0$                    &  $\neq 0$          \\
    NI             & $=0$                &  $\ne0$                   & $\ne 0$         &     $ = 0 $                     &  $\ne 0$   \\
    C              & $=0$                &  $=0$                     & $ =0 $          &     $ = 0 $                     &  $\ne 0$   \\
    SSI            & $=0$                &  $=0$                     & $\ne 0 $        &     $ = 0 $                     &  $ = 0$   \\
   \hline
    \end{tabular}
\label{tableI}
\end{table}

\begin{figure}
\includegraphics[trim = 10mm 0mm 0mm 0mm, clip=true, width=0.64\textwidth]{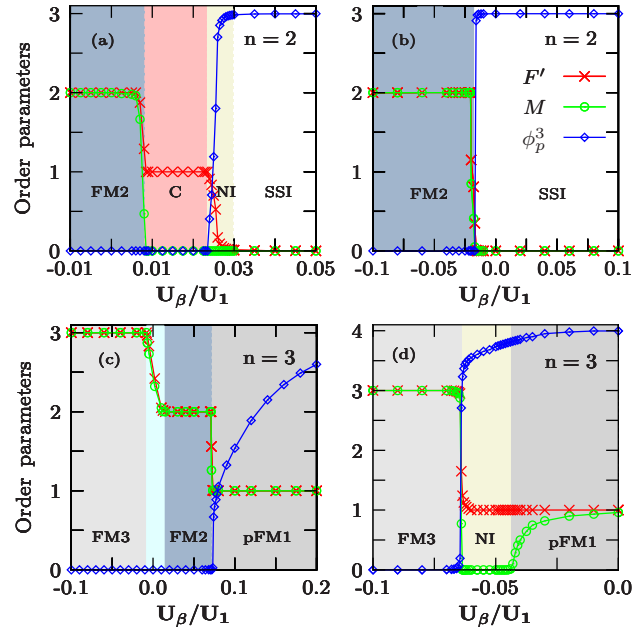}
\vspace{-8mm}
\caption{(Color online) Zero-temperature phase transitions for mixtures of spin-1 bosons in a 3D optical lattice with filling $n=2$ and $U_\gamma/U_1=0.025$ (a) and $-0.05$ (b) [Fig.~\ref{fig-fig2}(a)], and $n=3$ and $U_\gamma/U_1=0.2$ (c) and $-0.2$ (d) [Fig.~\ref{fig-fig2}(b)].}
\label{fig-fig3}
\end{figure}

{\it Phase diagrams from BDMFT}. The calculated magnetic phases from BDMFT are presented in Fig.~\ref{fig-fig2}. These phases are characterized by various on-site order parameters~\cite{Uniform}, including superfluid order $\phi^1_{\sigma m} = \langle b_{m\sigma}\rangle$, spontaneous magnetization $M = |\langle {\bf S}\rangle|$, spin magnitude $P = \langle {\bf S}^2\rangle$ (where ${\bf S} = {\bf S}_1 + {\bf S}_2$), nematic order $\phi^2_{\alpha\beta} = \langle S_\alpha S_\beta \rangle - {\delta_{\alpha\beta} \over 3} \langle S^2 \rangle$ and singlet pairing order $\phi_p^3 = \langle \Theta^\dagger\Theta\rangle$. The criteria for these different phases are summarized in Table~\ref{tableI}. These orders have also been adopted in other spinor BH models~\cite{spin1_Demler, spin2, spin22}. In these orders, the magnetization $M$ measures the spontaneous breaking of symmetry from a degenerate subspace to one of them with $M = F$. In the uniform phase without this symmetry breaking, $M = 0$. To characterize the possible mixing between different degenerate manifold, we define $P = F'(F'+1)$ (see Fig.~\ref{fig-fig3}, and $F' = F$ without mixing). The hidden high-order correlations between the multispin states should be detected by $\phi^2_{\alpha\beta}$ and $\phi^3_{p}$, where $\phi^3_{p}$ measures the pairing effect and $\phi^2_{\alpha\beta}$ the relative phase locking between spin components. We remark here that these order parameters are not orthogonal to each other, thus they may coexist in certain phases. Away from the Mott-insulating regime, the superfluid order $\phi^1_{\alpha m}$ is nonzero. With these order parameters, we characterize the whole phase diagrams for $n = 2$ and $n = 3$. 

\begin{figure}
\includegraphics[width=0.485\textwidth]{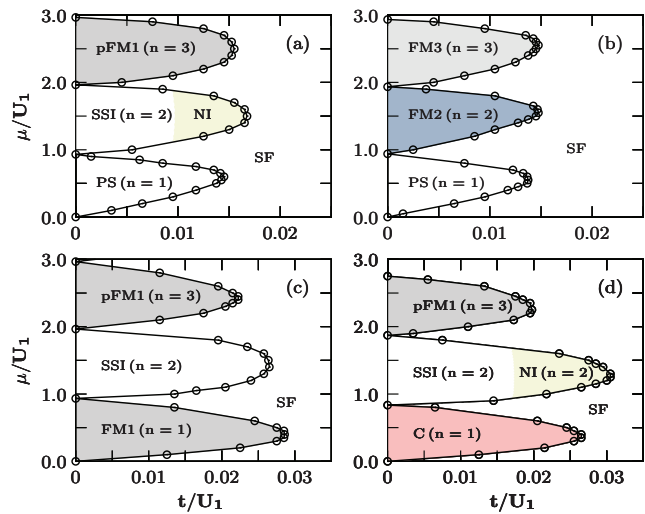}
\vspace{-7mm}
    \caption{(Color online) Mott insulator to superfluid transition of spin-1 heteronuclear atoms in a 3D optical lattice ($t \equiv t_1$). Parameters in (a),(c),(d) are $U_\beta = 0.032$ and $U_\gamma = 0.0011$ ($^{87}$Rb and $^{23}$Na), while in (b) are $U_\beta = -0.059$ and $U_\gamma =-0.002$. Other parameters are $U_2/U_1 = 1.92$, $t_2/t_1 = 3.78$, $U_\alpha/U_1 = 1.0$ (a),(b), $U_2/U_1 = 1.92$, $t_2/t_1 = 1.0$, $U_\alpha/U_1 = 1.0$ (c), and $U_2/U_1 = 1.0$, $t_2/t_1 = 1.0$, $U_\alpha/U_1 = 0.9$ (d).
    The phase separation phase in (a) and (b) is abbreviated as PS. }
    \label{fig-fig4}
\end{figure}

For $n = 2$ [Fig.~\ref{fig-fig2}(a)], we find five different competing phases in the $U_\gamma - U_\beta$ plane. According to Fig.~\ref{fig-fig1}, the SSI phase marks the regime when $F = 0$ and $g = 0$, thus the system is simply in the spin singlet insulating (SSI) phase. The spontaneous magnetization can be found when $F = 2$ and $g = 5$, which is denoted as FM2. The phase boundary between SSI phase and FM2 phase is well described by the change of degree of degeneracy at $3U_\gamma = U_\beta < 0$ [see Eq.~(\ref{Eq_2})]. The regime with odd parity ($F = 1$) is most intriguing, due to the possible existence of the cyclic (C) and the ferromagnetic (FM1) phase, which can be tuned by the interaction strengths, although all these phases are created from the same degenerate manifold. Between the C phase and SSI phase, we also find a narrow regime for nematic insulating (NI) phase. While the NI phase has been widely investigated in the spin-1 bosonic particles, we find that this phase is greatly suppressed in our model. Since in our simulations we have essentially considered an infinite system, the small regime for the NI phase should not be attributed to finite-size effects. The experimental regime to observe the NI phase will be discussed in more details in Fig.~\ref{fig-fig4}.

The order parameters as a function of interaction strengths are presented in Fig.~\ref{fig-fig3}(a) and (b) for $U_\gamma/U_1 = 0.025$ and $-0.05$ respectively, which represents attractive and repulsive interactions between the atoms. In the FM2 phase, we find the total magnetization $M = 2$. With the increasing of $U_\beta/U_1$, a transition from FM2 phase to FM1 (for $U_\gamma > 0$) and SSI phase (for $U_\gamma < 0$) is expected. In the former case we find the magnetization drops from 2 in FM2 phase to 1 in FM1 phase. In the SSI phase, we find that only the singlet pairing order is nonzero with $\phi^3_p = 3$ (see Ref.~\onlinecite{spin22}, and discussion in Ref. \onlinecite{Supp}). We find that, the NI phase appears in a small parameter regime between the C and SSI phases, as shown in Fig.~\ref{fig-fig3}(a), where the singlet pairing order and nematic order coexist simultaneously with vanishing magnetism. We remark here that the existence of the singlet pairing order ($\phi^3_p \neq 0$) is consistent with our single particle analysis, as shown in Ref.~\onlinecite{Supp}, and that nematic order $\phi^2_{\alpha\beta}$ is not presented in Fig.~\ref{fig-fig3} to simplify the figure, which is also nonzero for the NI phase.

The phase diagram for $n = 3$ is presented in Fig.~\ref{fig-fig2}(b), which is totally different from the phase diagram in Fig.~\ref{fig-fig2}(a). In the regime when $F = 3$ and $g = 7$ , we observe the spontaneous magnetization phase with $M = 3$ (denoted as FM3 phase), while in the regime with $F = 2$ and $g = 5$, we find the similar magnetization phase with $M =2$ for FM2. Between the FM3 and FM2 phase we find a broad mixed phase (MX) due to the coupling between the $g = 5$ and $g = 7$ degenerate manifolds with closed energies. The similar regime can also be found for $n = 2$ in Fig.~\ref{fig-fig2}(a), but this mixed regime is much smaller. Again, the most intriguing regime is for $F = 1$ and $g = 3$, in which the NI phase and paired FM phase, which is now denoted as pFM1 phase, can be realized. We can understand these phases from the evolution of order parameters as a function of $U_\beta/U_1$ in Fig.~\ref{fig-fig3}(c) and (d). In the NI phase, we find that the nematic order and singlet pairing order coexist, whereas all these three orders coexist in the pFM1 phase (see the materials in Ref.~\onlinecite{Supp}, where we prove that $\phi^3_p$ is nonzero only in the regime with $F = 1$ and $g = 3$).
Note here that we observe a second-order transition from the NI to the pFM1 phase.

The above interesting phases mainly occur in the regime for bosonic particles with odd parity, thus cannot been seen in identical bosonic particles. The whole phase diagrams for heteronuclear atoms are also totally different from the phase diagrams with identical atoms~\cite{spin2,spin22}. For instance, in two identical atoms, only the C phase can be found in the regime of the FM1 phase with on-site degeneracy $g=3$. For three identical atoms, the cyclic and trimer phases appear, instead of paired FM1 (pFM1) and anti-paralleling FM2 phases (FM2) in the parameter regime studied here. These observations highlight the unique features of heteronuclear mixtures.


{\it Mott to superfluid transitions}. We now ask the general question that how and where these phases can be found in experiments. Away from the deep Mott-insulating regime, which is frequently encountered in experiments, quantum fluctuations become more and more important with the increase of tunneling amplitudes; until finally the tunneling dominates in the superfluid regime with $\phi^1_{\alpha m} \neq 0$ (see definition in Table~\ref{tableI}, and magnetism of weakly interacting bosons can be found in Ref.~\onlinecite{hetero, hetero1, hetero2, hetero3, hetero4}). These fluctuation effects can be naturally included in our BDMFT approach.

Our calculated Mott to superfluid transition is presented in Fig.~\ref{fig-fig4} for different filling factors. The calculated diagrams depend strongly not only on the tunneling $t_m$, but also on the values of $U_{\alpha,\beta,\gamma}$ and $U_m'$. All these spin orders are stable against quantum fluctuations in the Mott-insulating lobes. Phase separation may be found in the $n = 1$ lobe when the two heteronuclear atoms have large difference in tunneling amplitudes and interaction strengths; otherwise, we will find the FM1 or C phase. In the $n =2$ lobe, we find the SSI, NI and FM2 phases in different parameter regimes. Especially, we find that it is possible to drive the SSI into the NI phase by tuning the tunneling amplitude [see Fig.~\ref{fig-fig4}(a),(d)]. While in Fig.~\ref{fig-fig2}(a), the NI phase can only be observed in a narrow parameter regime, here we find that this phase can be found in a wide parameter regime by controlling the system parameters. In the $n =3$
lobe, we find the pFM1 and FM3 phases, while the FM2 phase should be found in other system parameters. These observations demonstrate the experimental observability of the novel magnetic phases predicted in Fig.~\ref{fig-fig2}.

We finally discuss the experimental relevance of our theory. Recently, heteronuclear mixtures of spinor $^{23}$Na and $^{87}$Rb bosonic gases have been realized in an optical dipole trap~\cite{DJWang15}, and quantum phases of homonuclear spinor $^{23}$Na gases in optical lattices explored by overcoming the heating problem induced by the long-time thermalization~\cite{boson6,boson7}. For this reason, in Fig.~\ref{fig-fig4}(a),(c),(d) we have adopted the experimental parameters $U_\beta$ and $U_\gamma$ for these two atoms, and $U^2_\alpha<U_1 U_2$ to avoid phase separation. All parameters in the generalized BH model can be tuned independent, for example, the many-body interactions may be tuned via microwave~\cite{micro_resonance, micro_resonance1} or optical Feshbach resonances~\cite{optical_resonance, optical_resonance1, optical_resonance2, optical_resonance3, optical_resonance4, optical_resonance5, optical_resonance6, optical_resonance7}. The microscopic structure of these phases maybe detected using Bragg scattering~\cite{bragg} or optical birefringence~\cite{Natu,Stamper-Kurn}. The gapped spin-singlet insulator has a nonzero gap to all excitations, which can be measured by Bragg scattering, and ferromagnetic phases has a nonzero local spin polarized to a certain direction, which can be measured via spin-dependent light-atom interactions through dispersive birefringent imaging~\cite{Natu,Stamper-Kurn}. Recently, the spin nematic order in spinor gases was directly measured via a study of the magnetization noise after spin rotation~\cite{NI}.

To conclude, we show that for heteronuclear atoms, the angular momentum composition allows both even and odd parity states even for bosonic atoms, which can give rise to new exotic magnetic phases in the odd parity regimes. We address this issue via the bosonic dynamical mean-field theory approach and map out the complete phase diagrams as a function of many-body interaction strengths, focusing on the $n = 2$ and $n = 3$ Mott lobes. These phases are characterized by magnetization order, nematic order, singlet pairing order and spin magnitude order, which are not only determined by the on-site degeneracy, but also the competitions from many-body interactions. Their possible relevant regimes and parameters are also presented.

\textit{Acknowledgements.} We acknowledge useful discussions with Da-Jun Wang, Ying-Mei Liu, Ji-Ze Zhao and Zeng-Xiu Zhao. Y.L. is supported by the National Natural Science Foundation of China under Grants No. 11304386 and 11774428, and M.G. is supported by the National Youth Thousand Talents Program (No. KJ2030000001), the USTC start-up funding (No. KY2030000053).


\newpage
\begin{widetext}
\section{Supplementary Material}
\renewcommand{\theequation}{S\arabic{equation}}
\renewcommand{\thefigure}{S\arabic{figure}}
\renewcommand{\bibnumfmt}[1]{[S#1]}
\renewcommand{\citenumfont}[1]{S#1}
\setcounter{equation}{0}
\setcounter{figure}{0}

\section{Extended Bose-Hubbard model}
A spinor bosonic gas in an optical lattice is a mixture of hyperfine states of the same isotope. They system can undergo transitions between macroscopically occupied hyperfine states due to spin-exchange collisions, but it as a whole is in the ground state. For a system of bosonic gases with hyperfine spin $f$, for example, the spin-dependent interactions can be written in the second-quantized form:
\begin{eqnarray}
V({\bf x}_1-{\bf x}_2) = \frac{4\pi \hbar^2}{M_a} \sum^{2f}_{F=0} a_F \mathcal{P}_{F}  \delta({\bf x}_1- {\bf x}_2),
\end{eqnarray}
where $\mathcal{P}_{F}\equiv\sum_m \left|F,m_F\rangle \langle F, m_F \right|$ is the projection operator, $|F,m_F\rangle$ is the total hyperfine spin state formed by two atoms each with spin $f$, $M_a$ the atomic mass, and $a_F$ is the s-wave scattering length in the channel of total spin $F$. The spin-dependent interactions can also be written in the form of spin operators~\cite{Ho_98}. For example, for two identical spin-1 atoms, it can be expressed as:
\begin{eqnarray}
V({\bf x}_1-{\bf x}_2)  &=& (g_0\mathcal{P}_0 + g_2\mathcal{P}_2) \delta({\bf x}_1- {\bf x}_2) \nonumber\\
                        &=&(c_0 +  c_2 {\bf {S}}_1 \cdot {\bf {S}}_2)\delta({\bf x}_1- {\bf x}_2),
\end{eqnarray}
where $c_0=\frac{4\pi \hbar^2}{M_a} \frac{a_0+a_2}{3}$, $c_2=\frac{4\pi \hbar^2}{M_a} \frac{a_2-a_0}{3}$, and ${\bf S}_i$ is the spin operator of the $i$th
atom with spin-1. For two heteronuclear spin-1 atoms~\cite{Ham_two_spin}, however, the interactions take this form:
\begin{eqnarray}
V({\bf x}_1-{\bf x}_2) &=& (g_0\mathcal{P}_0 + g_1\mathcal{P}_1 + g_2\mathcal{P}_2) \delta({\bf x}_1- {\bf x}_2) \nonumber \\
                       &=& (\alpha + \beta {\bf S}_1 \cdot {\bf S}_2 + \gamma \mathcal{P}_0) \delta({\bf x}_1- {\bf x}_2)
\end{eqnarray}
with $\alpha = \frac{2\pi \hbar^2}{M_\mu} \frac{a_1+a_2}{2}$, $\beta=\frac{2\pi \hbar^2}{M_\mu} \frac{a_2-a_1}{2}$ and $\gamma=\frac{2\pi \hbar^2}{M_\mu}\frac{2a_0-3a_1+a_2}{2}$ with the reduced mass $M_\mu\equiv \frac {M_1M_2}{M_1+M_2}$ [$M_{1}$ ($M_2$) denotes the atomic mass for species 1 (2)].

Following the standard derivation for ultracold spinor gases, the many-body Hamiltonian for a system of heteronuclear mixtures of spin-1 condensate takes the following form~\cite{DJWang15_appen}:
\begin{eqnarray}\label{Ham_appen}
\hat{H} &=& \int d {\bf x} \hat{\Phi}^\dagger_{m\sigma} ({\bf x}) \left(-\frac{\hbar \Delta^2}{2M_{m\sigma}}+ V_{m\sigma}({\bf x})\right) \hat{\Phi}_{m\sigma} ({\bf x}) \nonumber \\
&+&   \int d {\bf x} \int d {\bf x^\prime} \Bigg[ \frac {c_{0m}}{2} \hat{\Phi}^\dagger_{m\sigma} ({\bf x}) \hat{\Phi}^\dagger_{m\sigma^\prime} ({\bf x^\prime})
\hat{\Phi}_{m\sigma^\prime} ({\bf x}^\prime) \hat{\Phi}_{m\sigma} ({\bf x}) + \frac{c_{2m}}{2} \hat{\Phi}^\dagger_{m\sigma} ({\bf x}) \hat{\Phi}^\dagger_{m\sigma^\prime} ({\bf x^\prime}) {\bf S}_{\sigma \sigma^{\prime\prime\prime}}\cdot {\bf S}_{\sigma^\prime \sigma^{\prime\prime}} \hat{\Phi}_{m\sigma^{\prime\prime}} ({\bf x}^\prime) \hat{\Phi}_{m\sigma^{\prime\prime\prime}} ({\bf x})\Bigg] \nonumber \\
&+&  \int d {\bf x} \int d {\bf x^\prime} \Bigg[ \alpha\hat{\Phi}^\dagger_{1\sigma}({\bf x}) \hat{\Phi}_{1\sigma}({\bf x}) \hat{\Phi}^\dagger_{2\sigma} ({\bf x}^\prime)  \hat{\Phi}_{2\sigma} ({\bf x}^\prime) + \beta \hat{\Phi}^\dagger_{1\sigma} ({\bf x}) {\bf S}_{\sigma \sigma^{\prime}} \hat{\Phi}_{1\sigma^\prime} ({\bf x}) \cdot \hat{\Phi}^\dagger_{2\sigma^{\prime\prime}} ({\bf x}^\prime) {\bf S}_{\sigma^{\prime\prime} \sigma^{\prime\prime\prime}}  \hat{\Phi}_{2\sigma^{\prime\prime\prime}} ({\bf x}^\prime)\Bigg] \nonumber \\
&+& \int d {\bf x} \int d {\bf x^\prime} \gamma \frac{(-1)^{\sigma-\sigma^\prime}}{3} \hat{\Phi}^\dagger_{1\sigma} ({\bf x}) \hat{\Phi}_{1\sigma^\prime} ({\bf x})
\hat{\Phi}^\dagger_{2-\sigma} ({\bf x}^\prime) \hat{\Phi}_{2-\sigma^\prime} ({\bf x}^\prime)
\end{eqnarray}
where $\hat{\Phi}_{m\sigma} ({\bf x})$ is the field annihilation operator for the $m$ species ($m=1,2$) in the hyperfine state $ | 1, \sigma \rangle$ at point ${\bf x}$.

We assume a deep optical lattice potential, and consider only the lowest energy band in the following. We also assume that the Wannier function of the lowest energy band $\omega_{m\sigma}({\bf x}-{\bf x}_i)$ is well localized in the $i$th lattice site. Expanding a field operator by Wannier functions of the lowest energy
band, $\hat{\Phi}_{m\sigma} ({\bf x}) = \sum_i b_{im\sigma} \omega_{m\sigma}({\bf x}-{\bf x}_i)$, Eq.(~\ref{Ham_appen}) reduces to a tight-binding
Bose-Hubbard model for heteronuclear mixtures of spin-1 bosons in an optical lattice, and the corresponding Bose-Hubbard model under the single-mode approximation~\cite{single_mode_98} can be written as:
\begin{eqnarray}\label{Ham2_appen}
    \hat{H} &=& -\sum_{\langle ij \rangle,m,\sigma} t_m ({b}^\dagger_{im\sigma}{b}_{jm\sigma} + {\rm H.c.}) - \sum_{i,m}\mu_m {n}_{i m}  \nonumber \\
    &+& \sum_{i,m} \bigg[\frac12U_{ m} {n}_{i m}(\hat{n}_{i m}-1) + \frac12U^\prime_{ m} ({\bf S}^2_{i m}-2{n}_{i m}) + U_\alpha n_{i1}n_{i2}  + U_\beta {\bf S}_{i1} \cdot {\bf S}_{i2} + \frac {1}{3} U_\gamma \Theta^\dagger_i \Theta_i\bigg],
\end{eqnarray}
where ${b}^\dagger_{im\sigma}$ (${b}_{i\sigma}$) is the bosonic creation (annihilation) operator of hyperfine state $m_F =\sigma$ for species $m=1,2$ at site $i$, ${n}_{im}=\sum_\sigma {n}_{im\sigma}$ with ${n}_{im\sigma}\equiv {b}^\dagger_{im\sigma}{b}_{im\sigma}$ being the number of particles, ${\bf S}_{im}\equiv {b}^\dagger_{im\sigma}{\bf \Gamma}_{\sigma\sigma^\prime}{b}_{im\sigma^\prime}$ is the local total spin operator (${\bf \Gamma}_{\sigma
\sigma^\prime}$ being the usual spin matrices for a spin-1 particle), $\Theta^\dagger \equiv {b}^\dagger_{i11}{b}^\dagger_{i-12} - {b}^\dagger_{i01}{b}^\dagger_{i02} + {b}^\dagger_{i-11}{b}^\dagger_{i12}$, $\mu_{m}$ denotes the chemical potential, and $t_{m}$ the hopping matrix element between nearest neighbors on the lattice. The third term in Eq.~(\ref{Ham2_appen}) describes a Hubbard repulsion with $U_{1,2}=(g^{(0)}_{1,2}+2g^{(2)}_{1,2})/3 \int d{\bf r} |\omega_{1,2}({\bf r} - {\bf r_i})|^4$, and the fourth term describes on-site spin-dependent interactions with $U^\prime_{1,2}=(g^{(2)}_{1,2}-g^{(0)}_{1,2})/3\int d{\bf r} |\omega_{1,2}({\bf r} - {\bf r_i})|^4$. Here, $g^{(s)}_{1,2}=4\pi \hbar^2 a^{(s)}_{1,2} /M_{1,2} $ with $a^{(s)}_{1,2}$ being the s-wave scattering length in the total spin $s=0$ and $2$ channels for species $1$ and $2$, $M_{1,2}$ the atomic mass, and $\omega_{1,2}({\bf r} - {\bf r_i})$ the Wannier function of the lowest energy band localized at the $i$th lattice site. The last three terms describe inter-species interactions with $U_{\alpha}=(g^{(1)}_{12}+g^{(2)}_{12})/2 \int d{\bf r} |\omega_{1}({\bf r} - {\bf r_i})|^2|\omega_{2}({\bf r} - {\bf r_i})|^2$, $U_{\beta}=(g^{(2)}_{12}-g^{(1)}_{12})/2 \int d{\bf r} |\omega_{1}({\bf r} - {\bf r_i})|^2|\omega_{2}({\bf r} - {\bf r_i})|^2$, and $U_{\gamma}=(2g^{(0)}_{12}-3g^{(1)}_{12} + g^{(2)}_{12})/2 \int d{\bf r} |\omega_{1}({\bf r} - {\bf r_i})|^2|\omega_{2}({\bf r} - {\bf r_i})|^2$, where $g^{(s)}_{12}=2\pi \hbar^2 a^{(s)}_{12} /M_{\mu} $ with $a^{(s)}_{12}$ being the s-wave scattering length in the total spin $s=0$, $1$ and $2$ channels.

\section{Wave functions for $n = 2$ and $n = 3$}

\subsection{Effective Hamiltonian, eigenvalues and eigenvectors}

   We first consider two heteronuclear atoms trapped in a single site, which is described by $H=U_{\alpha}n_1n_2+U_{\beta}{\bf S}_1 \cdot {\bf S}_2 + {1 \over 3} U_{\gamma}{\Theta}^{\dag}{\Theta}$,
    where $n_m = \sum_{s=-1}^{+1} b_{ms}^\dagger b_{ms}$. For non-identical atoms, we construct the basis as $|s; \sigma\rangle = b_{1s}^\dagger |0\rangle \otimes b_{2\sigma}^\dagger |0\rangle$,
    where $s,\sigma \in \{-1,0,1\}$, under which the above Hamiltonian can be expressed as,
\begin{align}
    H =
    \begin{pmatrix}
	U_{\alpha}+U_{\beta} & 0 & 0 & 0  & 0  & 0   & 0& 0  & 0 \\
	0 & U_{\alpha} & U_{\beta}  & 0  & 0  & 0   & 0  & 0 & 0   \\
	0 &U_{\beta} & U_{\alpha}  & 0  & 0  & 0   & 0  & 0  & 0  \\
	0 & 0 &0  &U_{\alpha}  & U_{\beta} & 0   & 0  & 0 & 0  \\
	0 & 0 & 0   & U_{\beta} &U_{\alpha} & 0   & 0  & 0  & 0   \\
	0 & 0 & 0   & 0  & 0 & U_{\alpha}+U_{\beta}   & 0  & 0  & 0  \\
	0 & 0 &0  & 0  & 0  & 0   & U_{\alpha}-U_{\beta}+\frac {U_{\gamma}} {3}  & \frac {U_{\gamma}} {3}  & U_{\beta}-\frac {U_{\gamma}} {3}  \\
	0 & 0 & 0   & 0  & 0 & 0   & \frac {U_{\gamma}} {3}  &  U_{\alpha}-U_{\beta}+\frac {U_{\gamma}} {3} & U_{\beta}-\frac {U_{\gamma}} {3}   \\
	0 & 0 & 0   & 0  & 0  & 0 & U_{\beta}-\frac {U_{\gamma}} {3}  & U_{\beta}-\frac {U_{\gamma}} {3}  &U_{\alpha}+\frac {U_{\gamma}} {3}
\end{pmatrix}.
\end{align}
The eigenvectors and eigenvalues for the above Hamiltonian can be decoupled into three sectors $F = 0, 1, 2$. We find for $F=2$,
$E_{F=2, g=5} = U_{\alpha}+U_{\beta}$, and the wave functions are
\begin{equation}
    \psi_{2, 2} = |1;1\rangle, \quad \psi_{2, 1}= |1;0\rangle + |0;1\rangle, \quad \psi_{2, 0} = \frac 12 |1;-1\rangle+\frac 12 |-1;1\rangle+|0;0\rangle, \quad \psi_{2, -1} = |-1;0\rangle + |0;-1\rangle\quad \psi_{2,-2} = |-1;-1\rangle.
\end{equation}
For $F=1$, $E_{F=1, g=3} = U_{\alpha}-U_{\beta}$, and we have
\begin{equation}
	\psi_{1,1} = |0;1\rangle - |1;0\rangle,\quad
    \psi_{1, 0} = |-1;1\rangle - |1;-1\rangle, \quad \psi_{1, -1} = |0;-1\rangle - |-1;0\rangle.
\end{equation}
For $F=0$, $E_{F=0, g=1} = U_{\alpha}-2U_{\beta}+U_\gamma$, the wave function is unique, and we have
\begin{equation}
    \psi_{0,0} = |0;0\rangle - |1;-1\rangle - |-1;1\rangle.
\end{equation}
Notice that the above wave functions are not normalized.

Next we consider the case with two identical atoms ($m =1$) and one heteronuclear atom ($m = 2$), therefore $n = 3$. The Hamiltonian can be written as
\begin{equation}
    H={1 \over 2} U_1 n_1(n_1-1)+ {1 \over 2} U_1^{'}({\bf S}_1^2-2n_1)+U_{\alpha}n_1n_2+U_{\beta} {\bf S}_1\cdot {\bf S}_2+ {1 \over 3} U_{\gamma}{\Theta}^{\dag}{\Theta}.
\end{equation}
We may represent the basis as $|s, s'; \sigma \rangle = \mathcal{A} b_{1s}^\dagger b_{1s'}^\dagger|0\rangle \otimes b_{2\sigma}^\dagger | 0\rangle$,
where $\mathcal{A}$ is the normalization constant. Moreover, we denote $|2_s;\sigma\rangle = |s,s;\sigma\rangle = (b_{1s}^\dagger)^2|0\rangle \otimes b_{2\sigma}^\dagger |0\rangle$.
Then the Hamiltonian can be represented as,
\begin{equation}
H=M_0\oplus M_1 \oplus M_{1} \oplus M_{-1} \oplus M_{2} \oplus M_{-2} \oplus M_{3} \oplus M_{-3},
\end{equation}
where
\begin{eqnarray}
    M_{3}  && = \begin{pmatrix}
	        U_1+{U_1}^{'}+2U_{\alpha}+2U_{\beta}
               \end{pmatrix}, \quad
    M_{-3}  = \begin{pmatrix}
	        U_1+{U_1}^{'}+2U_{\alpha}+2U_{\beta}
                \end{pmatrix}, \\
    M_{2} && = \begin{pmatrix}
	            U_1+{U_1}^{'}+2U_{\alpha}+U_{\beta}&\sqrt 2 U_{\beta}\\
	            \sqrt 2 U_{\beta}&U_1+{U_1}^{'}+2U_{\alpha}
                \end{pmatrix}, \quad
    M_{-2}  =\begin{pmatrix}
	            U_1+{U_1}^{'}+2U_{\alpha}+U_{\beta}&\sqrt 2 U_{\beta}\\
	            \sqrt 2 U_{\beta}&U_1+{U_1}^{'}+2U_{\alpha}
               \end{pmatrix}, \\
    M_{1} && = \begin{pmatrix}
	    U_1+2U_{\alpha} & \sqrt 2 {U_1}^{'} & \sqrt 2 U_{\beta} & 0 \\
	    \sqrt 2 {U_1}^{'} & U_1-{U_1}^{'}+2U_{\alpha}+\frac {U_{\gamma}} {3} & U_{\beta}-\frac {U_{\gamma}} {3} & \frac{\sqrt 2}{3} U_{\gamma} \\
	    \sqrt 2 U_{\beta}& U_{\beta}-\frac {U_{\gamma}} {3}& U_1+{U_1}^{'}+2U_{\alpha}+\frac {U_{\gamma}} {3} & \sqrt 2(U_{\beta}-\frac {U_{\gamma}} {3})\\
	    0 & \frac{\sqrt 2}{3}U_{\gamma} & \sqrt 2(U_{\beta}-\frac {U_{\gamma}} {3})& U_1+{U_1}^{'}+2U_{\alpha}-2U_{\beta}+\frac 23 U_{\gamma}\\
            \end{pmatrix}, \\
    M_{-1}&& =\begin{pmatrix}
	U_1+2U_{\alpha} & \sqrt 2 {U_1}^{'} & \sqrt 2 U_{\beta} & 0 \\
	\sqrt 2 {U_1}^{'} & U_1-{U_1}^{'}+2U_{\alpha}+\frac {U_{\gamma}} {3} & U_{\beta}-\frac {U_{\gamma}} {3} & \frac{\sqrt 2}{3} U_{\gamma} \\
	\sqrt 2 U_{\beta}& U_{\beta}-\frac {U_{\gamma}} {3}& U_1+{U_1}^{'}+2U_{\alpha}+\frac {U_{\gamma}} {3} & \sqrt 2(U_{\beta}-\frac {U_{\gamma}} {3})\\
	0 & \frac{\sqrt 2}{3}U_{\gamma} & \sqrt 2(U_{\beta}-\frac {U_{\gamma}} {3})& U_1+{U_1}^{'}+2U_{\alpha}-2U_{\beta}+\frac 23 U_{\gamma}
\end{pmatrix}, \\
    M_{0} && =\begin{pmatrix}
	U_1+{U_1}^{'}+2U_{\alpha}-U_{\beta}+\frac {U_{\gamma}} {3} & U_{\beta} & \sqrt 2(U_{\beta}-\frac {U_{\gamma}}{3}) & \frac {U_{\gamma}} {3} \\
	U_{\beta} & U_1-{U_1}^{'}+2U_{\alpha} & \sqrt 2 {U_1}^{'} & U_{\beta} \\
	\sqrt 2(U_{\beta}-\frac {U_{\gamma}} {3}) & \sqrt 2 {U_1}^{'} & U_1+2U_{\alpha}+\frac 23 U_{\gamma} & \sqrt 2(U_{\beta}-\frac {U_{\gamma}} {3})\\
	\frac {U_{\gamma}} {3} & U_{\beta} & \sqrt 2(U_{\beta}-\frac {U_{\gamma}} {3})& U_1+{U_1}^{'}+2U_{\alpha}-U_{\beta}+\frac {U_{\gamma}} {3}\\
\end{pmatrix}.
\end{eqnarray}
The above Hamiltonian can be decoupled into four sectors with $F = 1_\text{I}, 1_\text{II}, 2, 3$, with corresponding eigenvalues as
\begin{eqnarray}
    E_{F=3,g=7} && = U_1+{U_1}^{'}+2U_{\alpha}+2U_{\beta}, \quad E_{F=2, g=5} = U_1+{U_1}^{'}+2U_{\alpha}-U_{\beta},\\
    E_{F=1_\text{I}, g=3} && = \frac{1}{6}(6 U_1-3 {U_1}^{'}+12 U_{\alpha}-9 U_{\beta}+4 U_{\gamma}-\sqrt{81 {{U_1}^{'}}^2-162 {{U_1}^{'}} U_{\beta}+81 {U_{\beta}}^2+48 {{U_1}^{'}} U_{\gamma}-48 {U_{\beta}} U_{\gamma} +16 {U_{\gamma}}^2}). \nonumber
\end{eqnarray}
For $E_{F=3,g=7}$, we find the wave functions are
\begin{align}
	&\psi_{3, 3} = |2_1;1\rangle, \quad
	\psi_{3, 2}=\sqrt 2 |1,0;1\rangle+|2_1;0\rangle, \quad \psi_{3,1}=2|2_0;1\rangle+\sqrt{2}|1,-1;1\rangle+2\sqrt2 |1,0;0\rangle+|2_1;-1\rangle, \notag\\
	&\psi_{3,0}=|0,-1;1\rangle+|1,-1;0\rangle+\sqrt 2 |2_0;0\rangle+|1,0;-1\rangle,\quad
	\psi_{3,-1}=2|2_0;-1\rangle+\sqrt{2}|1,-1;-1\rangle+2\sqrt2 |0,-1;0\rangle+|2_{-1};1\rangle, \notag\\
	&\psi_{3, -2}=\sqrt 2 |0,-1;-1\rangle+|2_{-1};0\rangle, \quad
	\psi_{3, -3} = |2_{-1};-1\rangle.
\end{align}
For $E_{F=2,g=5}$, we find the wave functions are
\begin{align}
	&\psi_{2,2}=-\frac{1}{\sqrt 2}|1,0;1\rangle+|2_1;0\rangle, \quad
	\psi_{2,1}=-|2_0;1\rangle-\frac{1}{\sqrt 2}|1,-1;1\rangle+\frac{1}{\sqrt2}|1,0;0\rangle+|2_1;-1\rangle, \quad
	\psi_{2,0}=-|0,-1;1\rangle+|1,0;-1\rangle, \notag\\
	&\psi_{2,-1}=-|2_0;-1\rangle-\frac{1}{\sqrt 2}|1,-1;-1\rangle+\frac{1}{\sqrt2}|0,-1;0\rangle+|2_{-1};1\rangle, \quad
	\psi_{2,-2}=-\frac{1}{\sqrt 2}|0,-1;-1\rangle+|2_{-1};0\rangle.
\end{align}
For $F = 1$, the eigenvalues depends strongly on the value of $U_\gamma$, and so is their eigenvectors. These wave functions are too complex to be presented
here. However, when $U_\gamma=0$, they will take some simple form as following,
For $E_{F=1_\text{I},g=3}$, we find the wave functions are
\begin{equation}
	\psi_{1_{\text I},1}=-\frac{1}{\sqrt 2}|2_0;1\rangle+|1,-1;1\rangle, \quad
	\psi_{1_{\text I}, 0}=\sqrt2|1,-1;0\rangle+|2_0;0\rangle.\quad
	\psi_{1_{\text I},-1}=-\frac{1}{\sqrt 2}|2_0;-1\rangle+|1,-1;-1\rangle.
\end{equation}
For $E_{F=1_\text{II},g=3}$, we find the wave functions are
\begin{align}
&\psi_{1_{\text {II}},1}=\frac 13 |2_0;1\rangle+\frac{1}{3\sqrt2}|1,-1;1\rangle-\frac{1}{\sqrt 2}|1,0;0\rangle+|2_1;-1\rangle, \quad
\psi_{1_{\text {II}}, 0}=|0,-1;1\rangle-\frac 23 |1,-1;0\rangle-\frac{2\sqrt2}{3}|2_0,0\rangle+|1,0;-1\rangle, \notag \\
&\psi_{1_{\text {II}},-1}=\frac 13 |2_0;-1\rangle+\frac{1}{3\sqrt2}|1,-1;-1\rangle-\frac{1}{\sqrt 2}|0,-1;0\rangle+|2_{-1};1\rangle.
\end{align}

\subsection{Mean value of ${\Theta}^{\dag}{\Theta}$ for $\phi_p^3$}

With the above eigenvectors, we next calculate the expectation of operator ${\Theta}^{\dag}{\Theta}$, which is useful to understand the $\phi_p^3$
order in figures in the main text. In our following calculation, we re-express the operator as following,
\begin{align}
{\Theta}^{\dag}{\Theta}=&b_{11}^{\dag}b_{2-1}^{\dag}b_{11}b_{2-1}+b_{11}^{\dag}b_{2-1}^{\dag}b_{1-1}b_{21}-b_{11}^{\dag}b_{2-1}^{\dag}b_{10}b_{20} +b_{1-1}^{\dag}b_{21}^{\dag}b_{11}b_{2-1}+b_{1-1}^{\dag}b_{21}^{\dag}b_{1-1}b_{21}-b_{1-1}^{\dag}b_{21}^{\dag}b_{10}b_{20}\notag\\
&-b_{10}^{\dag}b_{20}^{\dag}b_{11}b_{2-1}-b_{10}^{\dag}b_{20}^{\dag}b_{1-1}b_{21}+b_{10}^{\dag}b_{20}^{\dag}b_{10}b_{20}
\end{align}

For $n = 2$, we find
\begin{align}
&{\Theta}^{\dag}{\Theta}|\psi_{2,M}\rangle=0\qquad M=-2, \cdots,2\notag\\
&{\Theta}^{\dag}{\Theta}|\psi_{1,M}\rangle=0\qquad M=-1, \cdots ,1\notag\\
&{\Theta}^{\dag}{\Theta}|\psi_{0, 0}\rangle=-3|1;-1\rangle-3|-1;1\rangle-3|0;0\rangle=3|\psi_{0,0}\rangle
\end{align}
Taken the normalization condition into account, we find
\begin{align}
&\langle\psi_{2,M}|{\Theta}^{\dag}{\Theta}|\psi_{2,M}\rangle=0\qquad M=-2, \cdots,2\notag\\
&\langle\psi_{1,M}|{\Theta}^{\dag}{\Theta}|\psi_{1,M}\rangle=0\qquad M=-1, \cdots,1\notag\\
&\langle\psi_{0,0}|{\Theta}^{\dag}{\Theta}|\psi_{0,0}\rangle=3.
\end{align}
The above results indicate that the order $\phi_p^3 =3 $ in the SSI phase regime, as shown in Fig. 3(b).

For $n = 3$, we find that
\begin{align}
&{\Theta}^{\dag}{\Theta}|\psi_{3,M}\rangle=0\qquad M=-3, \cdots,3\\
&{\Theta}^{\dag}{\Theta}|\psi_{2,M}\rangle=0\qquad M=-2, \cdots ,2.
\end{align}
Then
\begin{align}
&\langle\psi_{3,M}|{\Theta}^{\dag}{\Theta}|\psi_{3,M}\rangle=0\qquad M=-3, \cdots,3\\
&\langle\psi_{2,M}|{\Theta}^{\dag}{\Theta}|\psi_{2,M}\rangle=0\qquad M=-2, \cdots,2
\end{align}

We next focus on the wave functions at $U_\gamma = 0$. In the degenerate space of $1_{\text I}$,
\begin{align}
&{\Theta}^{\dag}{\Theta}|\psi_{1_{\text I},1}\rangle=|1,-1;1\rangle+\sqrt2|2_1;-1\rangle-|1,0;0\rangle \notag\\
&{\Theta}^{\dag}{\Theta}|\psi_{1_{\text I},0}\rangle=-\sqrt2|1,0;-1\rangle+2|2_0;0\rangle-\sqrt2 |0,-1;1\rangle\notag\\
&{\Theta}^{\dag}{\Theta}|\psi_{1_{\text I},-1}\rangle=|1,-1;-1\rangle+\sqrt2|2_{-1};1\rangle-|0,-1;0\rangle.
\end{align}
Taken the normalization condition into account, we find
\begin{equation}
\langle\psi_{1_{\text I},1}|{\Theta}^{\dag}{\Theta}|\psi_{1_{\text I},1}\rangle=\langle\psi_{1_{\text I},0}|{\Theta}^{\dag}{\Theta}|\psi_{1_{\text I},0}\rangle=\langle\psi_{1_{\text I},-1}|{\Theta}^{\dag}{\Theta}|\psi_{1_{\text I},-1}\rangle=\frac23
\end{equation}
In the second degenerate space of $1_{\text {II}}$,
\begin{align}
&{\Theta}^{\dag}{\Theta}|\psi_{1_{\text {II}},1}\rangle=\frac53 \sqrt2 |1,-1;1\rangle+\frac{10}{3}|2_{1},-1\rangle-\frac53 \sqrt2|1,0;0\rangle \notag\\
&{\Theta}^{\dag}{\Theta}|\psi_{1_{\text {II}}, 0}\rangle=\frac{10}{3}|1,0;-1\rangle-\frac{10}{3}\sqrt2|2_0;0\rangle+\frac{10}{3} |0,-1;1\rangle\notag\\
&{\Theta}^{\dag}{\Theta}|\psi_{1_{\text {II}},-1}\rangle=\frac53 \sqrt2 |1,-1;-1\rangle+\frac{10}{3}|2_{-1},1\rangle-\frac53 \sqrt2|0,-1;0\rangle
\end{align}
Taken the normalization condition into account, we find
\begin{equation}
\langle\psi_{1_{\text {II}},1}|{\Theta}^{\dag}{\Theta}|\psi_{1_{\text {II}},1}\rangle=\langle\psi_{1_{\text {II}},0}|{\Theta}^{\dag}{\Theta}|\psi_{1_{\text {II}},0}\rangle=\langle\psi_{1_{\text {II}},-1}|{\Theta}^{\dag}{\Theta}|\psi_{1_{\text {II}},-1}\rangle=\frac{10}{3}
\end{equation}
The above results demonstrate that the $\phi_p^3$ order is nonzero in the regime with $F = 1$, $g = 3$ for $n = 3$ in Fig. 1, which corresponds to the NI and pFM1 phases in Fig. 2. The value of $\phi_p^3 = 4$
observed in Fig. 3(d), which exceed $10/3$, can be realized by proper superposition of the three degenerate states.

\section{bosonic dynamical mean-field theory}
To investigate quantum phases of binary mixtures of spinor Bose gases loaded into a cubic optical lattice, described by Eq.~(\ref{Ham2_appen}), we establish a bosonic version of dynamical mean-field theory on the generic three-dimensional situation, and implement a parallel code to tackle the six-spin system with a huge Hilbert space. As in fermionic dynamical mean field theory, the main idea of the BDMFT approach is to map the quantum lattice problem with many degrees of freedom onto a single site - "impurity site" - coupled self-consistently to a noninteracting bath~\cite{georges96}. The dynamics at the impurity site can thus be thought of as the interaction (hybridization) of this site with the bath. Note here that this method is exact for infinite dimensions, and is a reasonable approximation for high but finite dimensions.

\subsection{BDMFT equations}
In deriving the effective action, we consider the limit of a high but finite dimensional optical lattice, and use the cavity method~\cite{georges96} to derive self-consistency equations within BDMFT. The effective action of the impurity site up to subleading order in $1/z$ is then expressed in the standard way~\cite{georges96, Byczuk_2008}, which is described by:
\begin{eqnarray}\label{eff_action}
S^{(0)}_\text{imp} &=& -\int_0^\beta \hspace{-0.2cm} d \tau d\tau' \sum_{\sigma\sigma'} \Bigg( \hspace{-0.1cm} \begin{array}{c} b^{(0)*}_{\sigma} (\tau)\quad b^{(0)}_{\sigma} (\tau) \end{array}\hspace{-0.1cm} \Bigg)^{\hspace{0.1cm}} \boldsymbol{\mathcal{G}}^{(0)-1}_{\sigma\sigma'}(\tau-\tau') \Bigg(\begin{array}{c} \hspace{-0.1cm} b^{(0)}_{\sigma'} (\tau')\\ b^{(0)*}_{\sigma'} (\tau') \end{array} \hspace{-0.1cm}\Bigg) 	\nonumber \\
&+&\int_0^\beta  d\tau\hspace{-0.1cm}  \left\{ \frac{1}{2}U_{ m}\; n^{(0)}_{ m}(\tau) \,\Big( n^{(0)}_{ m}(\tau) - 1 \Big)
+ \frac12U^\prime_{ m}\Big({\bf S}^{(0)}_{ m} (\tau)^2 - 2n^{(0)}_{ m}(\tau)\Big)- t_{ m} \sum_{\langle 0i\rangle,\sigma }
\Big( b^{(0)*}_{\sigma m}(\tau) \phi^{(0)}_{\sigma m}(\tau) + {\rm H.c.} \Big) \right\} \nonumber \\
&+& \int_0^\beta d\tau \left\{  U_{\alpha}n^{(0)}_{1} (\tau)  n^{(0)}_{2} (\tau) +
U_{\beta}{\bf S}^{(0)}_1 (\tau) \cdot {\bf S}^{(0)}_2 (\tau) +  \frac{1}{3} U_{\gamma} {\Theta}^{(0)*}_1 (\tau)  {\Theta}^{(0)}_2 (\tau) \right\}.\label{action}
\end{eqnarray}
Here, we have defined the Weiss Green's function (being a $12\times12$ matrix),
\begin{eqnarray}
&&\hspace{-5mm}\boldsymbol{\mathcal{G}}^{(0)-1}_{\sigma\sigma'}(\tau-\tau') \equiv - \\
&&\hspace{-5mm} \left(\begin{array}{cc} \hspace{-0.1cm}
				(\partial_{\tau'}-\mu_\sigma)\delta_{\sigma \sigma'}+ t^2
\hspace{-0.25cm} \sum \limits_{\langle 0i\rangle,\langle 0j\rangle} \hspace{-0.25cm}
G_{\sigma \sigma', ij}^1 (\tau, \tau')
			&  t^2  \hspace{-0.25cm} \sum \limits_{\langle
0i\rangle,\langle 0j\rangle} \hspace{-0.25cm}   G^2_{\sigma\sigma', ij}(\tau, \tau') \\
			 t^2  \hspace{-0.25cm} \sum \limits_{\langle
0i\rangle,\langle 0j\rangle} \hspace{-0.25cm} {G^2_{\sigma\sigma', ij}}^*(\tau', \tau)
	 & (-\partial_{\tau'}-\mu_\sigma)\delta_{\sigma \sigma'}+ t^2   \hspace{-0.25cm}
\sum \limits_{\langle 0i\rangle,\langle 0j\rangle} \hspace{-0.25cm} G_{\sigma \sigma', ij}^1
(\tau', \tau)
			 \hspace{-0.1cm} \end{array}\right) \hspace{-0.15cm},	\nonumber
\end{eqnarray}
and introduced
\begin{equation}
\phi^{}_{i,\sigma}(\tau) \equiv \langle b_{i, \sigma} (\tau)
\rangle_0
\end{equation}
as the superfluid order parameters, and
\begin{eqnarray}
\hspace{-0.5cm}G_{\sigma\sigma', ij}^1 (\tau, \tau')\hspace{-0.2cm}&\ \equiv\ &\hspace{-0.2cm}- \langle b_{i, \sigma} (\tau) b_{j, \sigma'}^* (\tau')
\rangle_0 + \phi_{i, \sigma'} (\tau) \phi_{j, \sigma}^* (\tau'), \\
\hspace{-0.5cm}G_{\sigma \sigma', ij}^2 (\tau, \tau')\hspace{-0.2cm}&\ \equiv\ &\hspace{-0.2cm}- \langle b_{i, \sigma} (\tau) b_{j, \sigma'} (\tau')
\rangle_0 + \phi_{i, \sigma'} (\tau) \phi_{j, \sigma} (\tau')
\end{eqnarray}
as the diagonal and off-diagonal parts of the connected Green's functions, respectively, where $\langle \ldots \rangle_0$ denotes the expectation value in the cavity system (without the impurity site). Note here that $\boldsymbol{\mathcal{G}}^{(0)-1}_{\sigma\sigma'}(\tau-\tau')$ is a $12\times12$ matrix with $\sigma$ ($\sigma^\prime$) running over all the possible values $-1$, $0$ and $1$ for the two species.

\subsection{Anderson impurity model}
The most difficult step in the procedure discussed above is to find a solver for the effective action. However, one cannot do this analytically. To obtain BDMFT equations, it is better to return back to the Hamiltonian representation. Here, the effective action, described by Eq. (2) in the main text, is represented by an Anderson impurity Hamiltonian
\begin{eqnarray}
\hat{H}^{(0)}_A &=& - \sum_\sigma  \Bigg( t_\sigma \Big(\phi^{(0)*}_{\sigma} \hat{b}^{(0)}_{\sigma} + {\rm H.c.} \Big) + \frac{1}{2}U_{\sigma} \hat{n}^{(0)}_{\sigma}\Big( \hat{n}^{(0)}_{\sigma} - 1\Big) + \frac12 U^\prime_{\sigma} \Big({\bf {\hat S}}^{(0)2}_{\sigma} - 2{\hat n}^{(0)}_{\sigma}\Big) - \mu_{\sigma} \hat{n}^{(0)}_{\sigma} \Bigg)  \\
         &+& U_{\alpha}{\hat n}^{(0)}_{1}  {\hat n}^{(0)}_{2} +
U_{\beta}{\bf {\hat S}}^{(0)}_1 \cdot {\bf {\hat S}}^{(0)}_2 +  \frac{1}{3} U_{\gamma} {\hat \Theta}^{(0)\dagger}_1  {\hat \Theta}^{(0)}_2
         + \sum_{l}  \epsilon_l \hat{a}^\dagger_l\hat{a}_l + \sum_{l,\sigma} \Big( V_{\sigma,l} \hat{a}^\dagger_l\hat{b}^{(0)}_{\sigma} + W_{\sigma,l} \hat{a}_l\hat{b}^{(0)}_{\sigma} + {\rm H.c.} \Big), \nonumber
\end{eqnarray}
where the chemical potential and interaction term are directly inherited from the Hubbard
Hamiltonian. The bath of condensed bosons is represented by the Gutzwiller term with
superfluid order parameters $\phi^{(0)}_{\sigma}$ for each component of the two species. The bath of normal bosons is
described by a finite number of orbitals with creation operators $\hat{a}^\dagger_l$ and energies
$\epsilon_l$, where these orbitals are coupled to the impurity via normal-hopping amplitudes
$V_{\sigma, l}$ and anomalous-hopping amplitudes $W_{\sigma, l}$. The anomalous hopping terms are
needed to generate the off-diagonal elements of the hybridization function.

We now turn to the solution of the impurity model. The Anderson Hamiltonian can straightforwardly be implemented in the Fock basis, and the corresponding solution can be achieved by exact diagonalization (ED) of fermionic DMFT~\cite{M. Caffarel_1994, georges96}. After diagonalization, the local Green's function, which includes all the information about the bath, can be obtained from the eigenstates and eigenenergies
in the Lehmann-representation
\begin{eqnarray}
G_{\rm imp,\sigma \sigma'}^1 (i \omega_n) &=& \frac{1}{Z} \sum_{mn} \langle m | \hat b_\sigma | n\rangle \langle n | \hat b_{\sigma'}^\dagger | m \rangle \frac{e^{- \beta E_n} - e^{-\beta E_m}}{E_n - E_m + i \hbar \omega_n} + \beta \phi_\sigma \phi^\ast_{\sigma'} \\
G_{\rm imp,\sigma \sigma'}^2 (i \omega_n) &=& \frac{1}{Z} \sum_{mn} \langle m | \hat b_\sigma | n\rangle \langle n | \hat b_{\sigma'} | m \rangle \frac{e^{- \beta E_n} - e^{-\beta E_m}}{E_n - E_m + i \hbar \omega_n} + \beta \phi_\sigma \phi_{\sigma'}.
\end{eqnarray}

Integrating out the orbitals leads to the same effective action as in Eq. (2) in the main text, if the following
identification is made
\begin{eqnarray}
\boldsymbol{\Delta}_{\sigma\sigma'} (i\omega_n)  & \equiv & t^2 {\sum_{\langle 0i\rangle ,\langle 0j \rangle }}^\prime\mathbf G^{(0)}_{\sigma\sigma',ij}(i\omega_n),
\end{eqnarray}
where $\sum^\prime$ means summation only over the nearest neighbors of the "impurity site", and we have defined the hybridization functions:
\begin{eqnarray}\label{hybridization}
\Delta_{\sigma\sigma'}^1(i \omega_n) & \equiv & -\sum_l\Big(\frac{V_{\sigma,l}V_{\sigma',l}}{\epsilon_l-i\omega_n} + \frac{W_{\sigma,l}W_{\sigma',l}}{\epsilon_l+i\omega_n}\Big) \nonumber \\
\Delta_{\sigma\sigma'}^2(i \omega_n) & \equiv &  -\sum_l\Big(\frac{V_{\sigma,l}W_{\sigma',l}}{\epsilon_l-i\omega_n} +
\frac{W_{\sigma,l}V_{\sigma',l}}{\epsilon_l+i\omega_n}\Big).
\end{eqnarray}
Hence, the Weiss Green's function can be expressed by the hybridization functions, and it reads
\begin{eqnarray}
\boldsymbol{\mathcal{G}}^{{(0)}-1}_{\sigma\sigma'}(i\omega_n) & = & (i\omega_n\sigma_z + \mu_\sigma)\delta_{\sigma\sigma'} -
\boldsymbol{\Delta}_{\sigma\sigma'} (i\omega_n) \nonumber \\
&=& \mathbf{\Sigma}_{\rm imp, \sigma\sigma'}(i\omega_n) + \mathbf{G}^{-1}_{\rm imp, \sigma\sigma'}(i\omega_n).
\label{ssG}
\end{eqnarray}

We make the approximation that the lattice self-energy $\Sigma_{\rm lat, \sigma\sigma'}$ coincides with the impurity self-energy $\Sigma_{\rm imp, \sigma\sigma'}$, and the self-consistency loop is then completed by the conditions for lattice Green's function
\begin{equation}
\mathbf{G}_{\rm lat}(\mathbf{k}, i\omega_n) = \frac 1 {i\omega_n\sigma_z + \mu_\sigma - \mathbf{\Sigma}_{\rm imp}(i\omega_n)- \epsilon_\mathbf{k}}\label{lattice_green}
\end{equation}
and for the superfluid order parameter
\begin{equation}
\phi^{(0)}_\sigma = \langle\hat{b}_\sigma\rangle_{0},
\label{ssphi}
\end{equation}
where the notation $\langle \ldots \rangle_0$ means that the expectation value is calculated in the cavity system~\cite{Walter}.
Equations (\ref{ssG}), (\ref{lattice_green}) and (\ref{ssphi}) thus consitute the set of BDMFT self-consistency conditions.


\subsection{Energy within BDMFT}
Results within BDMFT should not depend on the initial conditions of the self-consistency
loop. In some cases, however, the self-consistent BDMFT procedure yields multiple stable solutions, such as around the phase boundary of first-order transition. To find the ground state of the system in these cases, we need to compare energies of the coexisting solutions.
\subsubsection{Kinetic energy within BDMFT}
In order to calculate energy of the system, the starting point for our calculations is the Hamiltonian of spinor bosons in 3D optical lattice. Here, we only derive the formula of kinetic energy, since the energy on the impurity site straightforward for the Hubbard model. In terms of creation and annihilation operators for bosons, $b^\dagger_{i\sigma}$ and $b_{i\sigma}$, respectively, it has the form
\begin{equation}\label{Ham}
\hat{H}_{\rm kin} = - t\sum_{\langle ij \rangle,\sigma}(b^\dagger_{i\sigma}b_{j\sigma}  + b^\dagger_{j\sigma }b_{i\sigma}).
\end{equation}

We can express the creation and annihilation operators occurring in Eq.~(\ref{Ham}) in terms of operators that create and destroy particles in momentum states by the transformation
\begin{equation}
b_{i\sigma}= \frac{1}{V^{1/2}}\sum_{\rm p} {\rm e}^{-i {\bf p}\cdot {\bf R}_i/\hbar} a_{\sigma \rm p} =  \frac{1}{V^{1/2}(2\pi\hbar)^3}\int d{\bf p} {\rm e}^{-i {\bf p}\cdot {\bf R}_i/\hbar} a_{\sigma \rm p},
\end{equation}
with its reverse $a_{\sigma \rm p}=1/V^{1/2}\sum_{R_i} {\rm e}^{i{\bf p}\cdot{\bf R}_i}b_{i\sigma}$. Here, one assumes a Bose gas contained in a box of volume V. The Hamiltonian~(\ref{Ham}) then reads
\begin{eqnarray}
\hat{H}_{\rm kin}=\sum_{\sigma \rm p}\epsilon^0_{\rm p} a^\dagger_{\sigma \rm p} a_{\sigma \rm p},
\end{eqnarray}
with $\epsilon^0_{\rm p}$ being single-particle-state energy (or dispersion of the non-interacting tight-binding band $\epsilon^0_{ \rm p}=-\sum_{\langle ij \rangle} t_{ij}e^{i{\bf k}\cdot ({{\bf R_i}- {\bf R_j}})}$).

Then the expectation value of the kinetic energy operator can be given in this form:
\begin{eqnarray}
E_{\rm kin}=\sum_{\sigma}\int d {\bf p} \epsilon^0_{\rm p}\langle a^\dagger_{\sigma \rm p} a_{\sigma \rm p}\rangle = -k_BT\sum_{\sigma,n}\int d\epsilon\, \epsilon \rho(\epsilon) G_\sigma(i\omega_n),
\end{eqnarray}
where $i\omega_n=2n\pi/\beta$, and $\rho(\epsilon)$ denotes density of state with changing from momentum space $\bf p$ to energy space $\epsilon$. Here we have used the relation $G_{A,B}(\tau)=G_{B,A}(-\tau)$ for Green's function of bosons. Actually, we have this relation $G^\ast_\sigma(i\omega_n) = G_\sigma(-i\omega_n)$ for real parameters, and then the sum $\sum_n$ can just start from the positive part, i.e. $n\geq0$. Note here that there is a minus sign in the formula, in contrast to Fermi cases.

\subsubsection{Total energy of the impurity site}
The ground state within BDMFT corresponds to the solution with the lowest energy, where the total energy of the impurity site which is given as follows:
\begin{eqnarray}
E=E_{\rm kin}+E_{\rm int}.
\end{eqnarray}
For the BH model describing binary mixtures of spin-1 bosons in optical lattices, the on-site interaction term is give by:
\begin{eqnarray}
E_{\rm int}=\Bigg\langle \sum_\sigma \Big[\frac{1}{2}U_{\sigma} \hat{n}^{(0)}_{\sigma}( \hat{n}^{(0)}_{\sigma} - 1) + \frac12 U^\prime_{\sigma} ({\bf {\hat S}}^{(0)2}_{\sigma} - 2{\hat n}^{(0)}_{\sigma})\Big] + U_{\alpha}{\hat n}^{(0)}_{1}  {\hat n}^{(0)}_{2} +
U_{\beta}{\bf {\hat S}}^{(0)}_1 \cdot {\bf {\hat S}}^{(0)}_2 +  \frac{1}{3} U_{\gamma} {\hat{\Theta}}^{(0)\dagger}_1  {{\hat \Theta}}^{(0)}_2  \Bigg\rangle \nonumber
\end{eqnarray}

\end{widetext}

\begin{references}
\bibitem{Uniform} We have verified that all sites are uniform in the parameter regime studied here, and each order parameter in the whole lattice sites takes the identical value within our numerical accuracy. This assumption may break down in fermions (or bosons in other parameter regimes with $n=1$), due to the formation of antiferromagnetic phases, or in the models with gauge potentials, where the order parameters over more neighboring sites should be defined.
\end{references}
\end{document}